\documentstyle[psfig]{mn}

\newif\ifAMStwofonts

\ifoldfss
  \ifCUPmtlplainloaded \else
    \NewTextAlphabet{textbfit} {cmbxti10} {}
    \NewTextAlphabet{textbfss} {cmssbx10} {}
    \NewMathAlphabet{mathbfit} {cmbxti10} {} 
    \NewMathAlphabet{mathbfss} {cmssbx10} {} 
  \fi
  \ifAMStwofonts
    \ifCUPmtlplainloaded \else
      \NewSymbolFont{upmath} {eurm10}
      \NewSymbolFont{AMSa} {msam10}
      \NewMathSymbol{\upi}     {0}{upmath}{19}
      \NewMathSymbol{\umu}     {0}{upmath}{16}
      \NewMathSymbol{\upartial}{0}{upmath}{40}
      \NewMathSymbol{\leqslant}{3}{AMSa}{36}
      \NewMathSymbol{\geqslant}{3}{AMSa}{3E}

      \let\leq=\leqslant \let\leq=\leqslant
      \let\geq=\geqslant \let\geq=\geqslant
    \fi
  \fi
\fi 

\ifnfssone
  \newmathalphabet{\mathit}
  \addtoversion{normal}{\mathit}{cmr}{m}{it}
  \addtoversion{bold}{\mathit}{cmr}{bx}{it}
  \newmathalphabet{\mathbfit} 
  \addtoversion{normal}{\mathbfit}{cmr}{bx}{it}
  \addtoversion{bold}{\mathbfit}{cmr}{bx}{it}
  \newmathalphabet{\mathbfss} 
  \addtoversion{normal}{\mathbfss}{cmss}{bx}{n}
  \addtoversion{bold}{\mathbfss}{cmss}{bx}{n}
  \ifAMStwofonts
    \ifCUPmtlplainloaded \else
      \UseAMStwoboldmath
      \makeatletter
      \new@mathgroup\upmath@group
      \define@mathgroup\mv@normal\upmath@group{eur}{m}{n}
      \define@mathgroup\mv@bold\upmath@group{eur}{b}{n}
      \edef\UPM{\hexnumber\upmath@group}
      \new@mathgroup\amsa@group
      \define@mathgroup\mv@normal\amsa@group{msa}{m}{n}
      \define@mathgroup\mv@bold\amsa@group{msa}{m}{n}
      \edef\AMSa{\hexnumber\amsa@group}
      \makeatother
      \mathchardef\upi="0\UPM19
      \mathchardef\umu="0\UPM16
      \mathchardef\upartial="0\UPM40
      \mathchardef\leqslant="3\AMSa36
      \mathchardef\geqslant="3\AMSa3E

      \let\leq=\leqslant \let\leq=\leqslant
      \let\geq=\geqslant \let\geq=\geqslant
    \fi
  \fi
\fi 

\ifnfsstwo
  \DeclareMathAlphabet{\mathbfit}{OT1}{cmr}{bx}{it}
  \SetMathAlphabet\mathbfit{bold}{OT1}{cmr}{bx}{it}
  \DeclareMathAlphabet{\mathbfss}{OT1}{cmss}{bx}{n}
  \SetMathAlphabet\mathbfss{bold}{OT1}{cmss}{bx}{n}
  \ifAMStwofonts
    \ifCUPmtlplainloaded \else
      \DeclareSymbolFont{UPM}{U}{eur}{m}{n}
      \SetSymbolFont{UPM}{bold}{U}{eur}{b}{n}
      \DeclareSymbolFont{AMSa}{U}{msa}{m}{n}
      \DeclareMathSymbol{\upi}{0}{UPM}{"19}
      \DeclareMathSymbol{\umu}{0}{UPM}{"16}
      \DeclareMathSymbol{\upartial}{0}{UPM}{"40}
      \DeclareMathSymbol{\leqslant}{3}{AMSa}{"36}
      \DeclareMathSymbol{\geqslant}{3}{AMSa}{"3E}

      \let\leq=\leqslant \let\leq=\leqslant
      \let\geq=\geqslant \let\geq=\geqslant
    \fi
  \fi
\fi 

\ifCUPmtlplainloaded \else
  \ifAMStwofonts \else 
    \def\upi{\pi}
    \def\umu{\mu}
    \def\upartial{\partial}
  \fi
\fi

\title{A comparative analysis of empirical calibrators for nebular metallicity}
\author[E. P{\'e}rez-Montero \& A.I. D{\'\i}az]
       {Enrique~P{\'e}rez-Montero  \& Angeles I.~D{\'\i}az\\ 
        Departamento de F{\'\i}sica Te{\'o}rica, C-XI, Universidad Aut{\'o}noma de Madrid, 28049 Madrid, Spain\\}
        
\date{Accepted 
      Received ;
      in original form }

\pagerange{\pageref{firstpage}--\pageref{lastpage}}
\pubyear{2001}

\begin{document}

\maketitle

\label{firstpage}

\begin{abstract}

We present a new analysis of the main empirical
calibrators of oxygen abundance for ionised gas nebulae. With that aim we have compiled
an extensive sample of objects with emission line data including the near IR [SIII] lines
and the weak auroral lines which allow for the determination of the gas electron temperature. For all the 
objects the oxygen abundances have been derived in a homogeneus way, using the most 
recent sets of atomic coefficients and taking into the account the effect of particle
density on the temperature of O$^+$. The residuals between directly and empirically-derived 
abundances as a function of abundance have been studied. A grid of
photo-ionisation models, covering the range of physical properties of the gas, has been
used to explain the origin of the uncertainties affecting each abundance calibrator.
The range of validity for each abundance parameter has been identified and its average
uncertainty has been quantified.

\end{abstract}


\section{Introduction}

HII regions, from diffuse HII regions in the Galaxy to Extragalactic HII Regions 
(GEHR) as well as HII Galaxies, have been for many years, the main source of information
about metallicity in distant galaxies. Their bright emission line spectra are 
visible in all kinds of objects where there are recent episodes of star
formation. The analysis of these nebular spectra constitutes the best method, if not the
only one, for the determination of the chemical abundances of elements such as:
He, N, O, Ne, Ar, S having optical emission lines corresponding to different
ionization states. An accurate knowledge of these abundances is
essential for a complete understanding of the evolution of
stars and stellar systems and has allowed some light to be shed
on several  questions concerning the chemical evolution of galaxies in the Local 
Universe. They are now becoming even more relevant with the regard to the
younger Universe.

 Recombination lines yield the most accurate abundances because of their
weak dependence on nebular temperature. In fact, helium abundances can
be derived to an accuracy better than 5 \%.  Unfortunately, most of
the observed emission lines in ionised nebulae are collisionally
excited and their intensities depend exponentially on temperature. In
principle, this temperature can be determined from appropriate line
ratios, the most widely used being that of 
[OIII]$\lambda 4363$ {\AA}/$(\lambda 4959$ {\AA} $+ \lambda 5007$ {\AA}) although, recently, 
the improved sensitivity of new detectors in larger 
telescopes allows for the measurement of other auroral lines,
which are less temperature sensitive (Kinkel \&
Rosa 1994; Castellanos, D\'\i az \& Terlevich 2002; hereinafter CDT02). 

All these ratios involve the detection and
measurement of one intrinsically weak line which in many objects
is too faint to be observed. This is the case for regions with
high metal content -- where the efficient cooling exerted
by metallic ions renders weak lines undetectable --, HII
regions in distant galaxies and objects with low surface brightness. 
In these cases, empirical methods based on the intensities of strong,
easily observable, optical lines have been developed and are nowadays widely
used.  

The so called ``empirical methods'' are based on the cooling properties
of ionised nebulae which ultimately translate into a relationship between
emission line intensities and oxygen abundance. In fact, when the
cooling is dominated by oxygen, the electron temperature depends
inversely on oxygen abundance. Since the intensities of collisionally
excited lines depend exponentially on electron temperature, a relation is
expected to exist between these intensities and oxygen abundances.

According to Pagel et al. (1979), under the assumptions that a) the nebula is
ionisation bounded, b) the region can be represented by small clumps
of gas with a given electron density surrounded by a much less dense
material, so that the degree of ionisation is proportional to
($\epsilon^2 n_e Q_H)^{1/3}$ where $n_e$ is the clump electron density, $Q_H$ is the
number of hydrogen ionising photons and $\epsilon$ is the filling factor and c)
the cooling is fixed by oxygen abundance, we can consider that the
emission line spectrum of the nebula will depend on: the energy
distribution of the ionising radiation field, the ionisation parameter
and the oxygen abundance . Therefore, if a single relation
between the chosen calibrator and the oxygen abundance is sought,
further assumptions are needed implying that either the hardness of the
radiation field or the degree of ionisation or both depend on oxygen
abundance.

Following these ideas, several abundance calibrators have been proposed
involving different emission line ratios: among others, [OIII] $\lambda$ 5007 \AA/H$\beta$ (Jensen,
Strom \& Strom 1976; [OIII] $\lambda$ 5007 \AA/ [NII] $\lambda$ 6584
\AA(Alloin et
al. 1979) and ([OII] $\lambda$ 3727 \AA{} + [OIII] $\lambda$ 5007 \AA)/ H$\beta$ (R$_{23}$;
Pagel et al. 1979) . The advantages and drawbacks of the different calibrators have 
been discussed by several authors (see Pagel, Edmunds \& Smith 1980; 
Kennicutt \& Garnett 1996; Kewly \& Dopita 2002). Although 
abundances derived through the use of these calibrations are recognised to suffer from
considerable uncertainties, they are still believed by many authors to trace
large-scale trends in galaxies. 
Empirical methods have been used to derive
abundances in objects as different as dwarf irregular galaxies ({\it
  e.g.} Skillman, Kennicutt \& Hodge 1989), individual HII regions in
spiral galaxies ({\it e.g.} Oey \& Kennicutt 1993), low surface
brightness galaxies (McGaugh 1994), nuclear starbursts ({\it e.g.}
Storchi-Bergmann, Calzetti \& Kinney 1994) and even active galactic
nuclei (Storchi-Bergmann et al. 1998). They have also been employed in the
derivation of abundance distributions in the discs of spiral galaxies
({\it e.g.} Belley \& Roy 1992; van Zee et al. 1998) and emission line
galaxies at intermediate redshift ({\it e.g.} Kobulnicky \& Kewley 2004)

In this work we perform a comparative analysis of the principal empirical 
calibrations of abundances which are based on the 
intensities of the nebular lines of oxygen, nitrogen and sulphur, 
visible in the optical and far 
red spectral regions. All these calibrations present a considerable 
scatter, usually larger than that associated with observational errors
and probably related to the assumptions mentioned above.
The aim of our work is to understand the reasons for this scatter and, 
whenever possible, to find ways of improving the empirical derivation of abundances.

In order to do that, we have compiled a large sample of emission line objects (HII 
galaxies, GEHR and diffuse HII regions in the Galaxy and the Local Group) with a direct 
determination of the total oxygen abundance through the measurement of the auroral lines of 
[OIII]$\lambda$4363 \AA, [OII]$\lambda$ 7327 \AA\  or [SIII]$\lambda$6312 \AA\ and we have constructed 
a complete sequence of photo-ionisation models, covering the main 
physical properties of these objects. 

In the next section we describe the sample of objects and the process of determination of the
oxygen abundance. In Section 3 , we summarize the main properties of
the photoionisation models used for our analysis which is then presented in
Section 4. Finally, Section 5 summarizes our results and the main conclusions reached.

\begin{figure}
\begin{minipage}{85mm}
\psfig{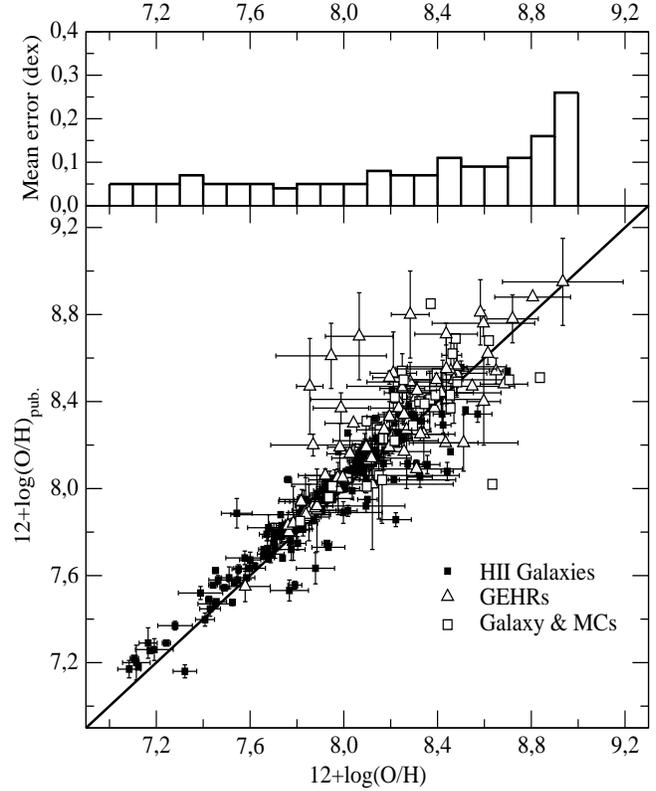}
\caption{Comparison, for the whole sample, between the oxygen abundances published in the
original sources and the abundances as calculated in this work. Symbols for this and subsequent 
plots are: solid squares for HII Galaxies, upward triangles for Giant Extragalactic
HII regions and open squares for Diffuse HII regions in the Galaxy and the Magellanic Clouds. 
The upper panel in the Figure shows the distribution of the observational errors (half-error bars) 
with oxygen abundance.}
\label{metpub}
\end{minipage}
\end{figure} 

\begin{figure}
\begin{minipage}{85mm}
\psfig{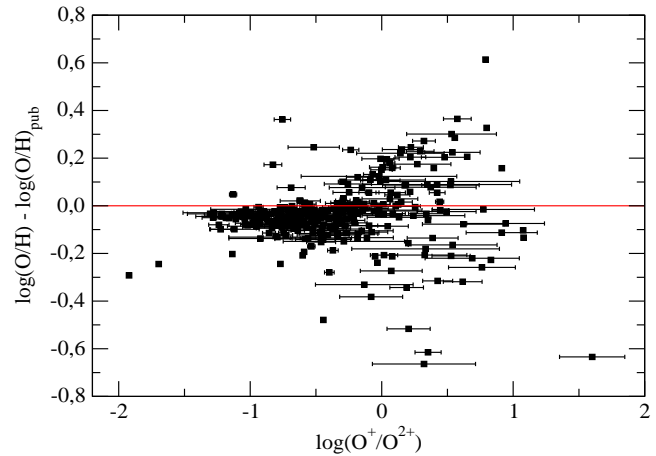}
\caption{Difference between the oxygen abundances calculated in this work and the published abundances
as a function of the $(O^+/O^{2+})$ ratio. Differences are much larger in
the low excitation regime, where the abundance of $O^+$ has more weight.}
\label{ionis}
\end{minipage}
\end{figure} 


\begin{table*}
\begin{minipage}{170mm}
\vspace{-0.3cm}
\normalsize
\caption{Bibliographic references for the emission line fluxes of the compiled sample}
\begin{center}
\begin{tabular}{lcc}
\hline
\hline
Reference\footnote{Those references marked with an $^*$, give fluxes of the emission lines of 
[SIII] in the near IR} & Object type\footnote{GEHR denotes Giant Extragalactic HII Regions, 
HIIG, HII Galaxies and  DRH, Diffuse HII Regions} & N\'{u}mber of points \\
\hline

Castellanos et al., 2002$^*$ & GEHR & 5 \\
Dennefeld \& Stasi\'{n}ska$^*$, 1983  & DHRs & 22 \\
D\'{\i}az et al., 1987$^*$ & GEHRs in NGC604 & 6 \\
Dinerstein \& Shields, 1986 & GEHRs & 2 \\
Edmunds \& Pagel, 1984  & GEHRs in NGC7793 & 3 \\
French, 1980            & HIIGs        & 13 \\
Garnett \& Kennicutt, 1994$^*$ & GEHR in M101 & 1 \\
Garnett et al., 2004$^*$ & GEHRs in M51 & 2 \\
Garnett et al., 1997$^*$ & GEHRs in NGC2403 & 11 \\
Gonz\'alez-Delgado et al., 1994$^*$ & GEHR in NGC2363 & 13 \\ 
Gonz\'alez-Delgado et al., 1995$^*$ & GEHRs in NGC7714 & 5 \\
Guseva et al., 2000 & HIIGs & 4 \\
Kennicutt, Bresolin \& Garnett, 2003$^*$  & GEHRs in M101 & 19 \\
Kinkel \& Rosa, 1994$^*$ & S5 in M101 & 1 \\
Kniazev et al., 2001  & HIIGs & 2 \\
Kunth \& Sargent, 1983  & HIIGs         & 13 \\
Kwitter \& Aller, 1981  & GEHRs in M33 &   5 \\
Izotov, Thuan \& Lipovetsky, 1994 & HIIGs & 10 \\
Izotov, Thuan \& Lipovetsky, 1997 & HIIGs & 27 \\
Izotov \& Thuan, 1998 & HIIG & 18 \\
Lequeux et al., 1979          & HIIG         & 8 \\
Pagel et al., 1979      & NGC300,7  & 1 \\
Pagel et al. 1992$^*$    & HIIGs & 10 \\
Pastoriza et al., 1993$^*$ & GEHRs in NGC3310 & 5 \\
Peimbert et al., 1986   & NGC2363 & 1 \\
Popescu \& Hopp, 2000 & HIIGs & 22 \\
Rayo et al., 1982        & GEHRs in M101 & 3 \\
Shaver et al., 1983      & DHRs & 7 \\
Shields \& Searle, 1978$^*$ & GEHRs in M101 & 2 \\ 
Skillman, C\^{o}t\'e \& Miller, 2003 & GEHRs in Sculptor & 6 \\ 
Skillman \& Kennicutt, 1993$^*$ & IZw18 & 2 \\
Skillman et al., 1994$^*$ & UGC4483 & 1 \\
Terlevich et al., 1991   & HIIGs & 100 \\
V\'{\i}lchez et al., 1988$^*$ & GEHRs in M33 & 5 \\ 
V\'{\i}lchez \& Esteban, 1996$^*$  & DHRs & 3 \\
V\'{\i}lchez \& Iglesias-P\'aramo, 2003 & GEHRs in Virgo & 9 \\
\hline
\end{tabular}
\end{center}
\label{refs}
\end{minipage}
\end{table*}


\section{Sample of objects and abundance derivation}

Our sample is comprised of a combination of different emission line
objects ionised by young massive stars: diffuse HII regions in the Galaxy and the 
Local Group (DHR), Giant Extragalactic HII regions (GEHR) and HII
galaxies (HIIG) and therefore does not include planetary nebulae or 
objects with non-thermal activity. For all of them direct determinations of electron
temperature exist thus allowing the derivation of the oxygen abundance
which we have taken as the observational metallicity indicator. The
sample includes the objects analysed in (D\'\i az \& P\'erez-Montero 2000,
hereinafter: DPM00) with the addition of low
excitation GEHRs from CDT02; GEHRs in M101 (Kennicutt,
Bresolin \& Garnett 2003) and M51 (Garnett, Kennicutt \& Bresolin 2004);
GEHRs in galaxies in the Sculptor Group
(Skillman, C\^{o}t\'e \& Miller 2003) and the Virgo cluster 
(V\'{\i}lchez \& Iglesias-P\'aramo 2004);  and HII galaxies from
the works of Guseva et al. (2000), Popescu \& Hopp (2000)and  Kniazev et al. (2001).
Data from these latter objects have been complemented with information in the spectral range between
7000 {\AA} and 1 micron from P\'erez-Montero \& D\'\i az (2003; hereinafter PMD03, 12 HII galaxies) 
and Garnett (1992; 13 objects), including the [SIII] strong emission lines.



\begin{table*}
\begin{minipage}{170mm}
\vspace{-0.3cm}
\normalsize
\caption{Recalculated electron densities of [SII], electron temperatures of
[OII] and [OIII] and ionic abundances of O$^+$ and O$^{2+}$ for the whole sample.}

\begin{center}
\begin{tabular}{ccc c c c c c c}
\hline
\hline
Object   & Region & Ref. & n([SII]) & t([OIII]) & 12+$\log\left(\frac{O^{2+}}{H^+} \right)$ & t([OII]) 
& 12+$\log\left(\frac{O^+}{H^+} \right)$ &  $\log\left(\frac{O}{H} \right)$ \\
\hline
NGC628   &  H13  & CDT02 &   64$\pm$42  & 1.00$\pm$0.04 & 7.70$\pm$0.03 & --          & 8.04$\pm$0.10 & 8.20$\pm$0.08 \\
NGC1232  &  CDT1 & CDT02 &  121$\pm$35  & 0.54$\pm$0.09 & 8.09$\pm$0.11 & --          & 8.86$\pm$0.27 & 8.93$\pm$0.25 \\
         &  CDT2 & CDT02 &  20:         & 1.15$\pm$0.19 & 7.45$\pm$0.10 & --          & 7.77$\pm$0.28 & 7.94$\pm$0.23 \\
         &  CDT3 & CDT02 &  198$\pm$61  & 0.82$\pm$0.08 & 7.76$\pm$0.07 & --          & 8.38$\pm$0.23 & 8.48$\pm$0.21 \\
         &  CDT4 & CDT02 &  224:        & 0.93$\pm$0.06 & 7.69$\pm$0.04 & --          & 8.08$\pm$0.14 & 8.23$\pm$0.12 \\
\hline
\end{tabular}

\end{center}

\label{data}
\end{minipage}
\end{table*}


The sources for the line intensities, together with the number and
class of the collected objects, are summarized in Table \ref{refs}.  
The total sample comprises 367 objects with lines in the optical part
of the spectrum, 282 of them with [NII] data, and 126 with near IR
[SIII] data.

The physical conditions -- electron temperature, electron density and
oxygen abundance -- for the whole sample have been recalculated using the same procedures 
as in PMD03, based on the fivel-level statistical equilibrium model in the task TEMDEN 
contained in the software package IRAF (De Robertis, Dufour \& Hunt 1987;
Shaw \& Dufour 1995). The atomic coefficients used are the same as in
PMD03 and are referenced in Table 4 of that work. Electron densities were determined from
the [SII] $\lambda$ 6717{\AA} / $\lambda$ 6731{\AA} line ratio. Electron temperatures
have been calculated  from the [OIII] ($\lambda$ 4959{\AA}+$\lambda$5007{\AA})/
$\lambda$ 4363{\AA} line ratio for all but
13 objects of the sample for which the [SIII] ($\lambda$
9069{\AA}+$\lambda$9532{\AA})/$\lambda$ 6312{\AA} line ratio has been used isnstead. These latter objects are of low excitation and lie on the high metallicity
range (for example,  CDT1 in NGC1232 (Castellanos
et al. 2002) or S5 in M101 (Kinkel \& Rosa, 1994)).
For them, an empirical relation between [OIII] and [SIII]
electron temperatures has been used:
\[t([OIII]) = 0.95t([SIII]) + 0.08\]
\noindent based on the grids of photo-ionisation models described in the next section and, differing
slightly from the empirical relation found by Garnett (1992), 
due mainly to the introduction of the new atomic coefficients
for S$^{2+}$ from Tayal \& Gupta (1999).  

Regarding [OII] temperatures, for 81 objects of
the sample it has been possible to
derive its value from the [OII]($\lambda$ 3726{\AA}+$\lambda$3729{\AA})
/$\lambda$ 7325{\AA} line ratio. \begin{footnote}{The [OII] $\lambda$7319{\AA}+$\lambda$7330{\AA} lines can have a contribution 
by direct recombination which increases with temperature. Using
the calculated [OIII] electron temperatures, we have estimated these contributions  to be less than 4 \% in all cases
and therefore we have not corrected for this effect}\end{footnote}. For the rest of the objects in the sample we have resorted to 
the model relations between 
t([OII]) and t([OIII]) found in PMD03 that take explicitly  into account the dependence of t([OII])
on electron density. This can affect the deduced abundances of $O^+/H^+$ 
by non-negligible factors, larger in all cases than the reported
errors.  

Figure \ref{metpub}, shows a comparison of the total oxygen abundances derived as described 
above with the values published in the original sources. The  deviations from the 1:1 relation
arise mostly in the high metallicity range as a result of the
dependence of t([OII]) on density which affects the calculated O$^+$/H$^+$
abundances. This can be better seen in Figure \ref{ionis} where the abundance
differences are plotted as a function of the O$^+$/O$^{++}$ ionic
fraction. 

The oxygen abundances of the sample objects cover 
the range 0.02Z$_\odot$ (IZw18; Skillman \& Kennicutt 1993) to 1.82Z$_\odot$ (CDT1 in NGC1232; 
Castellanos et al. 2002)\begin{footnote}
{A solar value of 12+log(O/H)= 8.69 (Allende-Prieto, Lambert \& Asplund
2001) is assumed through this paper.}
\end{footnote}. The re-calculated [SII] electron densities,
[OII] and  [OIII] electron temperatures
and oxygen abundances are listed  in Table \ref{data}. The quoted  uncertainties have been derived from the
emission line flux errors as published in the corresponding references. In the upper panel of Figure \ref{metpub}, 
the values of these errors (half-error bars) are plotted as a function of oxygen abundance. It can be seen that the 
errors are almost constant, with an average value of about $\pm$0.07 dex, up to 12+log(O/H)$\simeq$ 8.1 and thereafter  
increase with metallicity up to $\pm$ 0.5 dex at the highest derived abundances. 

The complete table will be
available in electronic form at CDS ({\em Centre de Donn\'ees astronomiques de Strasbourg}), 
via anonymous ftp to cdsarc.u-strasbg.fr (130.79.128.5),
(http://cdsweb.u-strasbg.fr) , or at
http://pollux.ft.uam.es/enrique/Table1/. Only an example is given here. 

At any rate, it should be recalled that the  determination of the  gaseous chemical abundances 
is usually accomplished by combining results from photoionization models  and
observed emission line intensity ratios. Even when the electron temperature can be determined with good accuracy, there are several major unsolved problems that severely limit the confidence of present
results including: (1) the effect of temperature structure in multiple-zone models (PMD03); 
(2) the presence of temperature fluctuations across 
a given nebula (Peimbert, 2003);  (3) collisional and density effects on ion temperatures (Luridiana, Peimbert, \& Leitherer 1999); 
(4) the presence of neutral zones affecting the determination of ionization correction factors 
(ICF's)(Peimbert, Peimbert \& Luridiana 2002); (5) the ionization structure which is not adequately reproduced by current
models (PMD03); (6) the possible photon leakage that affects the
low ionization lines formed in the outer parts of the ionized regions (Castellanos, D\'{\i}az, \& Tenorio-Tagle 2002). 
The first three effects can introduce uncertainties with respect to the derived oxygen abundances of about 0.2, 0.3 and 0.4 dex respectively,  depending on excitation. 
The uncertainties introduced by the latter effects have not yet been quantified.

\section{Photo-ionisation models}
In order to identify and understand the possible sources of scatter in the different empirical
calibrations, we have calculated a set of photo-ionisation models covering the physical 
conditions of the observed objects. No attempt has been made however to perform recalibrations using these models.
 
The photoionisation models have been computed with the most recent version of the 
photo-ionisation code CLOUDY 96 (Ferland 2002). 

Each modelled HII region has been assumed to be spherically 
symmetric, with the ionised emitting gas taken to be of constant
density (10 and 100 particles per cm$^3$), located at a 
distance very large compared to its thickness and 
therefore allowing the approximation of plane-parallel geometry. The gas is ionised by
a single massive star whose spectral energy distribution (SED) is represented by a CoStar NLTE  
stellar atmosphere (Schaerer \& de Koter, 1997) with effective temperature between 35000 K and 
50000 K . The impact  on the analysis of the general trends 
shown by empirical calibrations using other SEDs is negligible and, in practice, only slightly affects
the absolute effective temperature scale. 
Ionisation parameters ($U$)  between
10$^{-2}$ and 10$^{-3}$, which is the range corresponding to the
ionisation degree shown by the sample objects, have been chosen. 
Finally, the solar chemical abundances used are those given in Table \ref{abun}.

\begin{table}
\begin{minipage}{85mm}
\caption{Adopted solar abundances (
$^1$: Grevesse \& Sauval, 1998, $^2$: Holweger, 2001 and $^3$:
Allende-Prieto et al. 2001) with and without depletion factors 
for the refractory elements.}
\begin{center}
\begin{tabular}{ccc}
\hline
\hline
Element & Photosphere\footnote{in terms of $\log(X/H)$} & Depleted$^a$\\
\hline
He$^1$   &           -1.07  &          -1.07\\
O$^3$    &           -3.31   &        -3.31 \\
N$^2$    &           -4.07   &        -4.07 \\
S$^1$    &           -4.67   &        -4.67 \\
C$^2$    &           -3.41   &        -3.41 \\
Ne$^1$   &           -3.92   &        -3.92 \\
Ar$^1$   &           -5.60   &        -5.60 \\
Si$^2$   &           -4.47   &        -4.77 \\
Fe$^2$   &           -4.55   &        -5.55 \\
Mg$^1$   &           -4.47   &        -5.47 \\
Al$^1$   &           -5.53   &        -6.53 \\
Ca$^1$   &           -5.64   &        -6.64 \\
Na$^1$   &           -5.67   &        -6.67 \\
Ni$^1$   &           -5.75   &        -6.75 \\
\hline
\end{tabular}
\end{center}
\label{abun}
\end{minipage}
\end{table}

We have computed models with values of this solar abundance multiplied
by factors: 1.7, 0.85, 0.34, 0.17, 0.08 and 0.04 corresponding to
1, 0.5, 0.2, 0.1, 0.05 and 0.025 times the solar Grevesse \& Sauval (1998) value
($\log(O/H)$=-3.08). 
The refractory elements: Fe, Mg, Al, Ca, Na and 
Ni have been depleted by a factor of 10 and Si by a factor of 2 
(Garnett et al. 1995), to take into account 
the presence of dust grains.  In the case of nitrogen, we have considered, for the 
models with $U$ = 10$^{-2.5}$, another set of abundances with a value
of (N/O) 0.5 dex lower than the solar value, close to the values found in low metallicity nebulae.

For the sake of clarity,
only the models for $n_e$=100 cm$^{-3}$, T$_{eff}$ = 35000 K and 50000 K, and
$U$ = 10$^{-2.0}$ and 10$^{-3.0}$ are shown in the figures. 

\begin{figure*}
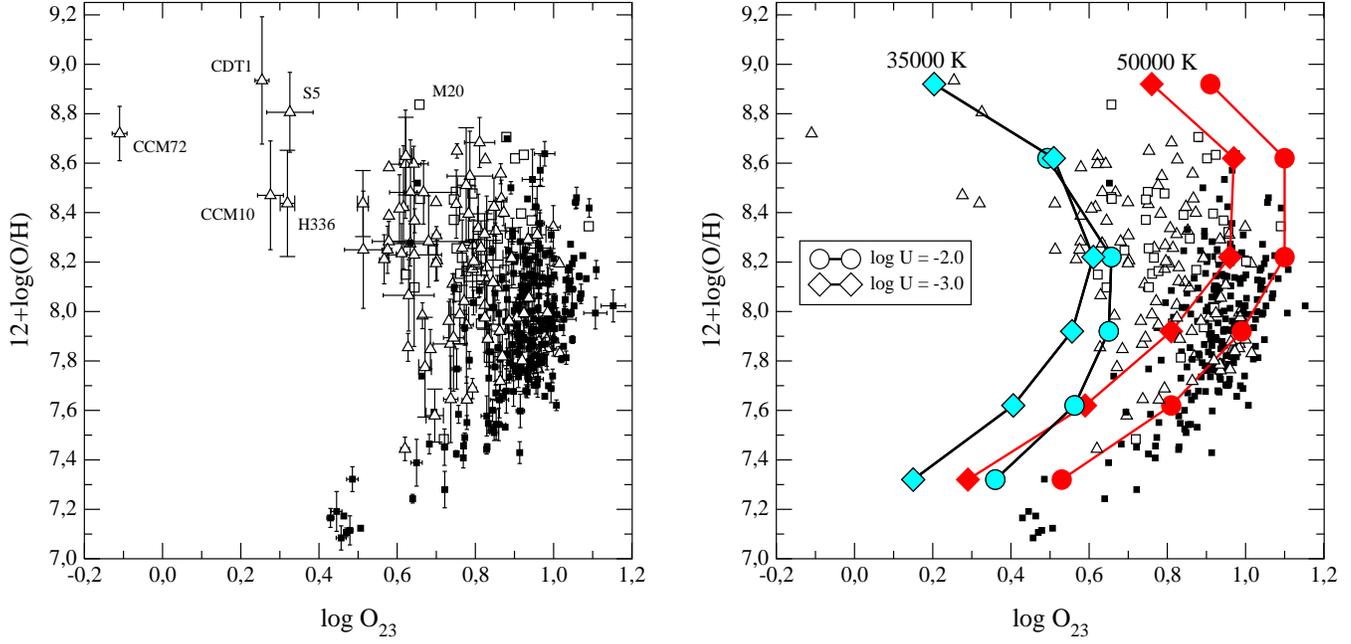

\begin{minipage}{175mm}
\centerline{
\psfig{figure=fig03l.eps,height=8.5cm,width=8.5cm,clip=}
\hspace{0.5cm}
\psfig{figure=fig03r.eps,height=8.5cm,width=8.5cm,clip=}}
\caption{Relation between O$_{23}$ and the metallicity, represented by 12+log(O/H), (left) 
and the comparison with CLOUDY photo-ionisation models (right) for 
different values of the effective temperature (35000 K in light tone and 50000 K in dark tone), 
metallicity (from 0.08 to 1.6Z$_{\odot}$) 
and ionisation parameter ($\log U$ = -2.0 (circles) and -3.0 (diamonds).}
\label{o23_o}
\end{minipage}
\end{figure*} 
\begin{figure*}
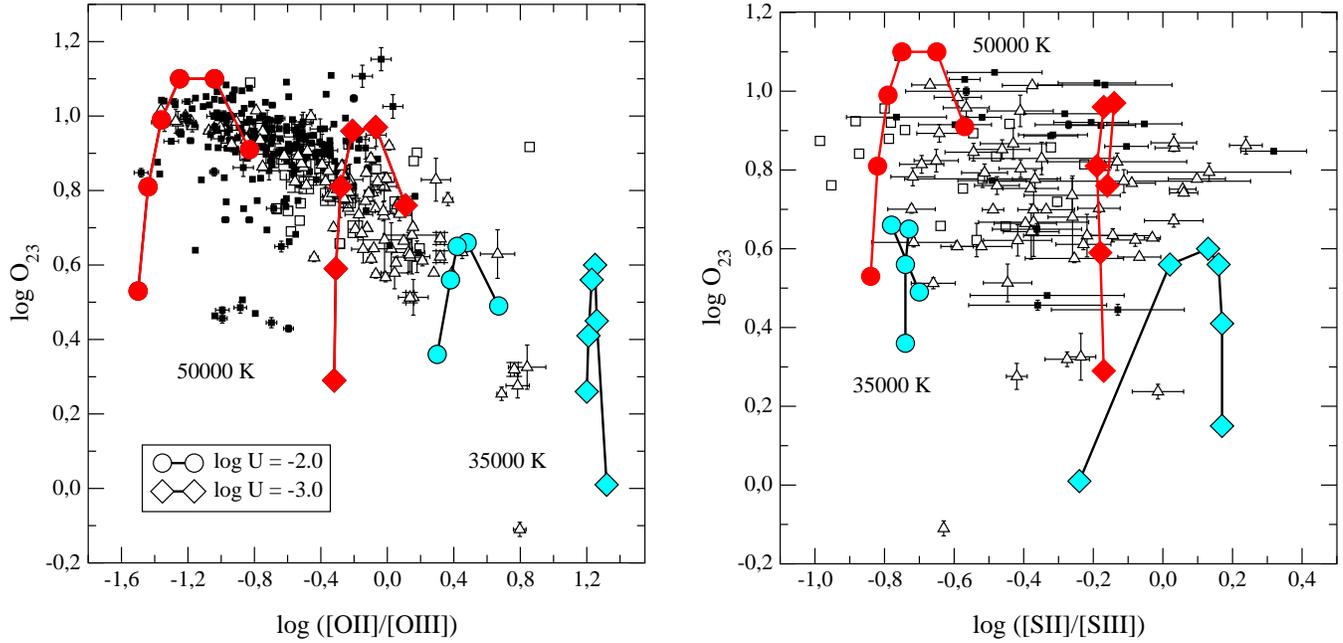

\begin{minipage}{175mm}
\centerline{
\psfig{figure=fig04l.eps,height=8.5cm,width=8.5cm,clip=}
\hspace{0.5cm}
\psfig{figure=fig04r.eps,height=8.5cm,width=8.5cm,clip=}}
\caption{Relation between the O$_{23}$ parameter and the ([OII]/[OIII]) ratio, which depends both on
effective temperature and ionisation parameter, (left) and the ([SII]/[SIII]) ratio, that dependes mostly on
ionisation parameter (right). Symbols for the models are the same as in Figure 3.}
\label{o23_2}
\end{minipage}
\end{figure*} 

\section{Empirical abundance parameters}

\subsection{The R$_{23}$ (O$_{23}$) parameter}

The R$_{23}$ parameter, that here we have preferred to rename as
O$_{23}$ in order to differentiate it from 
the analogous parameter based on sulphur emission lines, was defined as:\\
\[O_{23} \equiv \frac {{\mathrm I(3727\AA)+I(4959\AA)+I(5007\AA)}}{{\mathrm I(H}\beta)}\]

\noindent by Pagel et al. (1979). Its relation with oxygen abundance for the objects of the compiled sample can be seen 
in the left panel of Figure \ref{o23_o}. The relation  
is double-valued. This is due to 
the efficiency of oxygen as a cooling agent thus
decreasing the strength of the oxygen emission lines at high metallicities.  At low metallicities
however, the cooling is mainly exerted by hydrogen 
and the oxygen line strengths increase with metallicity. The value of logO$_{23}$ reaches a
maximum of about 1.2 at an oxygen abundance of 12+log(O/H) $\approx$ 
8.0. 

Three different regions can be distinguished in the plot: a lower
branch in which O$_{23}$ increases with incresing abundance,
an upper branch in which the opposite occurs and a turnover region. The two branches can, in principle, be fitted by regression lines
with positive and negative slope respectively providing a low to
moderate uncertainty in the determination of the metallicity. In the turnover 
region with 
log O$_{23}$ $\geq$ 0.8 and 12+log(O/H) 
$\geq$ 8.0, although the precise values are difficult to assess, objects showing the same value of O$_{23}$ can have oxygen
abundances that differ by almost an order of magnitude. It should be
noted that a large proportion of the data lie on top of this ill
defined zone (up to 40\% of the total number of objects  and even more
in the case of HII galaxies) where the abundance determination can be
very uncertain.

Another characteristic of the calibration which is readily apparent is the existence of a
scatter larger than accounted for by observational errors. This scatter
is related to the fact that, in general,  ionised regions do not constitute a single
parameter family, hence different geometries  of the
emitting gas (ionisation parameter) and different ionising radiation
temperatures can affect the values of the abundance parameter
O$_{23}$. 
This can be seen in the right panel of Figure \ref{o23_o}, where data are
shown together with different model sequences . Error bars have been omitted for the sake
of clarity. O$_{23}$ is seen to depend on both ionisation parameter and
stellar effective temperature. The dependence on ionisation parameter
is more evident at low metallicities, while
the dependence on effective temperature is important in all metallicity
regimes.  This double dependence makes of O$_{23}$ a rather unsuitable abundance
parameter. In fact, at an
oxygen abundance of 12+log(O/H) = 7.8, logO$_{23}$ can vary between 0.5
and 0.9. 

Different assumptions about the effects of metallicity
on either nebular structure or ionising temperature have been used by
different authors in order to define a sequence of models that would
eventually allow the calibration of the upper branch where
observational data is very scarce. From analyses of HII region data,
McCall, Rybski \& Shields (1985) concluded that the stellar ionising
temperature varied with metallicity while the filling factor remained
constant, whereas Dopita \& Evans (1986) concluded the opposite:
that ionising temperature was constant while U varied with oxygen
abundance. These two different assumptions led to calibrations yielding
abundances that differ by more than a factor of two.

\begin{figure}
\begin{minipage}{85mm}
\psfig{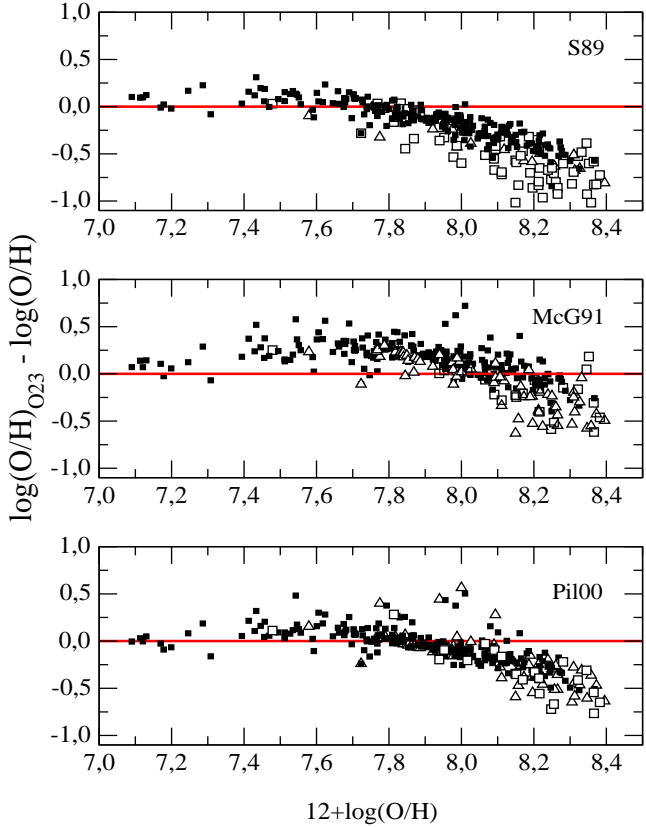}
\caption{Residuals of the metallicities deduced from the Skillman (1989; upper panel), McGaugh 
(1991; middle panel) and Pilyugin (2000; lower panel) calibrations as a function of the 
directly derived abundances for the lower branch of the O$_{23}$ vs. O/H plot.}  
\label{reslow}
\end{minipage}
\end{figure} 

\begin{figure}
\begin{minipage}{85mm}
\psfig{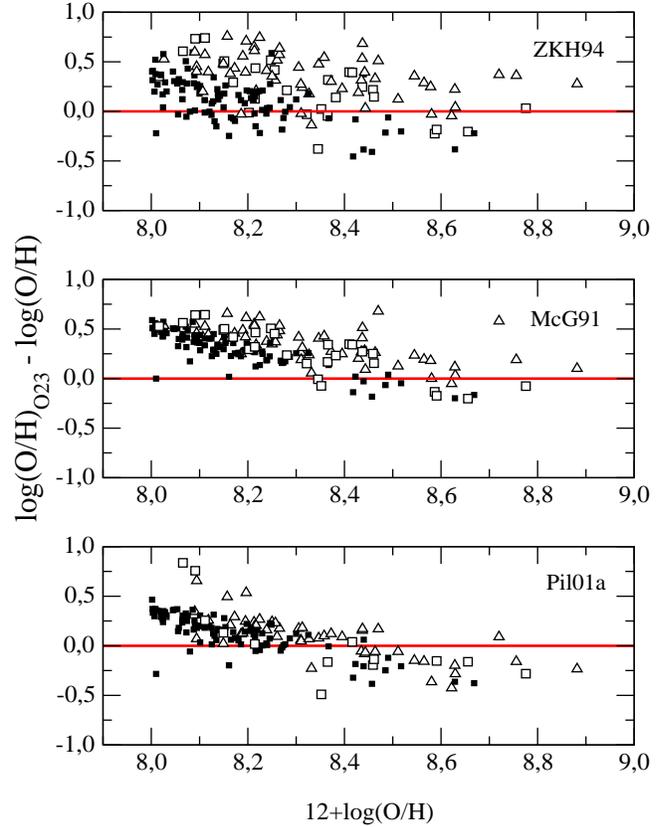}
\caption{Residuals of the metallicities deduced from the ZKH94 (upper panel), McGaugh 
(1991; middle panel) and Pilyugin (2001a; lower panel) calibrations as a function of the 
directly derived abundances for the upper branch of the O$_{23}$ vs. O/H plot.}  
\label{resup}
\end{minipage}
\end{figure} 

Theoretical stellar evolution models point to a relation between
stellar metallicity and effective temperature in the sense that, for a
given mass, stars of higher metallicities show lower effective
temperatures. This fact led McGaugh (1991;McG91) to produce a new calibration based
on more realistic theoretical models in which the ionisation is
provided by stellar clusters of different metallicities. According to
his models, in the upper branch, O$_{23}$ is relatively insensitive to
both ionising temperature and U and the models converge to a single
sequence. In the lower branch however, O$_{23}$ is mostly dependent on
U, as has already been shown by Skillman (1989; S89) and additional information 
about this parameter is needed in order to apply the empirical method with greater confidence. 

Some authors have used the [OII]/[OIII] ratio as an ionisation parameter
indicator to obtain this additional information ({\it e.g.} Kobulnicky et
al. 1999). The hardness of the ionising radiation however also affects this ratio in a
significant way. In fact, at
a given value of U, the [OII]/[OIII] ratio is
lower for higher stellar effective temperatures as a result of the
increase of the ionisation of O$^+$ to O$^{++}$.
These effects can be seen in  Figure \ref{o23_2} where $\log O_{23}$ is plotted as a function
of log ([OII]/[OIII]) (left panel) where it can be seen that the
[OII]/[OIII] ratios corresponding to models
with the same ionisation parameter and different stellar ionising
temperature widely differ.  

Just the opposite
happens in the case of [SII]/[SIII], another line ratio used as an
ionisation parameter indicator ({\it e.g} D\'\i az et al. 1991). At
constant U, [SII]/[SIII]  increases somewhat with increasing
stellar effective temperature as more S$^{++}$ is converted to
S$^{3+}$,  although in this latter case the effect is important only for
the highest ionisation parameters (U$\geq$ 10$^{-2}$), thus making the
[SII]/[SIII] ratio a more useful ionisation parameter
diagnostic. Figure \ref{o23_2} (right panel) shows that [SII]/[SIII]
depends mostly on the ionisation parameter and is rather insensitive to
the stellar effective temperature. 

Regarding observational data, no clear relation is found between $\log O_{23}$ and 
log ([SII]/[SIII]), implying that oxygen abundance and the ionisation
parameter are not correlated. On the other hand, a definite trend between
$\log O_{23}$ and [OII]/[OIII] is clearly seen which can be explained
by the expected dependence of stellar effective temperature and
metallicity (see also Kewley \& Dopita 2002).

\begin{figure*}
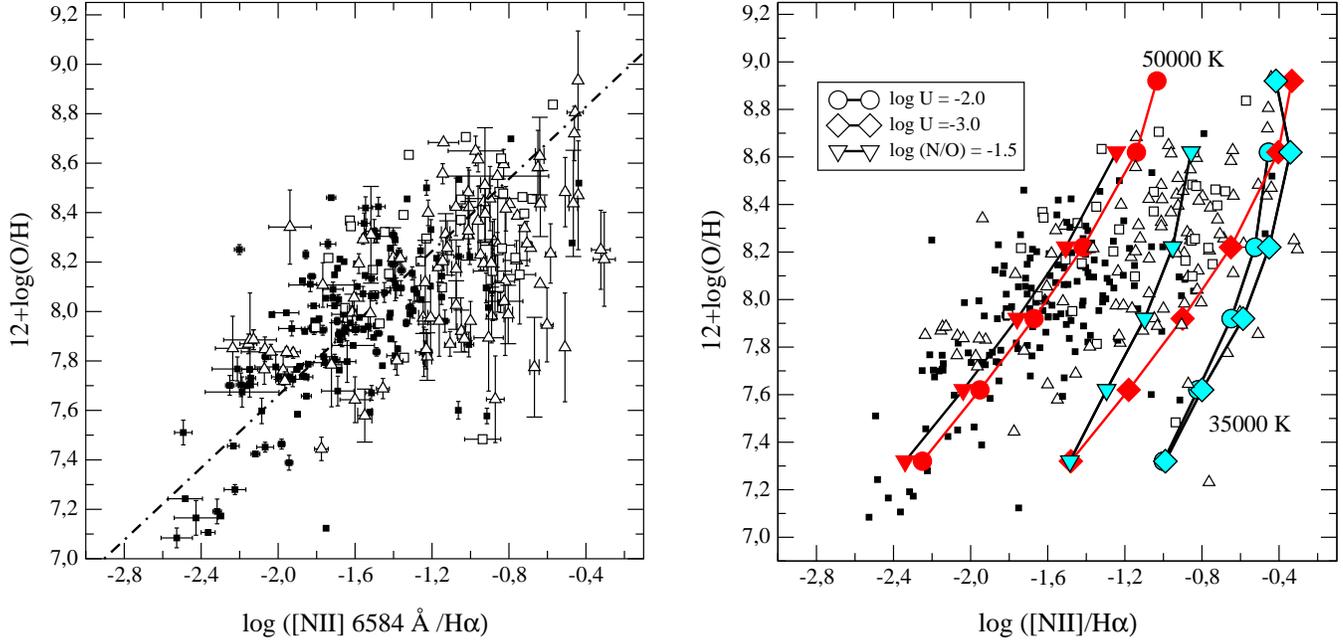

\begin{minipage}{175mm}
\centerline{
\psfig{figure=fig07l.eps,height=8.5cm,width=8.5cm,clip=}
\hspace{0.5cm}
\psfig{figure=fig07r.eps,height=8.5cm,width=8.5cm,clip=}}
\caption{Relation between N2 and 12+log(O/H) (at left) 
and the comparison with CLOUDY photo-ionisation models (at right) for 
different values of effective temperature (35000 K in light tone and 50000 K in dark tone), 
metallicity (from 0.08 to 1.6Z$_{\odot}$) 
and ionisation parameter ($\log U$ = -2.0 (circles) and -3.0 (diamonds).}
\label{N2_o}
\end{minipage}
\end{figure*} 

Finally, in the turnover region, O$_{23}$ is
sensitive to both ionisation parameter and ionising temperature and
is almost insensitive to oxygen abundance. 

It should be taken into account that McGaugh models use zero age
star clusters. The situation becomes much more complicated when the evolution
of these clusters is taken into account; the evolution of massive stars
is fast and metallicity dependent and the cluster ionising temperature
may not be a monotonicaly decreasing function of age once WR stars
begin to appear (Garc\'\i a-Vargas, Bressan \& D\'\i az 1995; Stasi\'nska \&
Leitherer 1996). 

Figures \ref{reslow} and \ref{resup}, show the residuals of several
published calibrations of the O$_{23}$ parameter in both lower and upper 
branches as well as in the intermediate region. The calibrations of S89 for the lower
branch and Zaritsky, Kennicutt \& Huchra (1994; hereinafter ZKH94)
for the upper one, involve only the O$_{23}$ parameter.  In contrast, the 
calibrations of McG91 and Pilyugin (2000, 2001a; hereinafter Pil00 and Pil01a,
respectively) take also into account the dependence on 
ionisation parameter and effective temperature 
via the ([OII]/[OIII]) ratio (Kobulnicky et al. 1999) and the parameter, P, defined as
the quotient of [OIII] and ([OII]+[OIII]) by Pil00. 

For the lower branch, that we have considered as corresponding to 12+log(O/H) $<$ 8.0, 
the best fit is found for the calibration by Pil00, with a mean value for the
metallicity 0.03 dex higher
than the mean value of those directly derived. Its uncertainty in this 
regime, understood as the standard deviation of the residuals, is
$\pm$0.15 dex which should be compared to the $\pm$0.07 dex average error in the
direct oxygen abundance determination (see Figure \ref{metpub}).
The calibrations by Skillman and McGaugh present also the same dispersion, 
although the mean values of the deduced abundances are 0.06 dex lower and 0.21 dex higher
respectively than the mean value of those directly derived. A slight
trend in the sense of abundances being more underestimated as the
metallicity increases is found in the S89 calibration. This trend is probably
introduced by the slight dependence of O$_{23}$ on ionisation
parameter at high effective temperature that as shown by HII galaxy
data in left panel of Figure  \ref{o23_2}.

In fact, in this metallicity range, he $O_{23}$ values predicted by photoionisation models 
for different values of the ionisation parameter and stellar effective temperature produce 
a scatter in the $O_{23}$ versus 12+log(O/H) relationship larger than shown by observational data. 
This probably indicates that the objects compiled
to perform the calibrations, mainly HII galaxies, show very similar
properties, {\it \i.e.} they show a very restricted range of ionisation
parameters and ionising temperatures. It should be noted however, that
not all HII galaxies share those properties and that, in particular,
those that lack detectable electron temperature sensitive lines and
therefore are the probable targets of the empirical calibrations, show
ionisation parameters which are significantely lower. These include
Luminous Compact Blue Galaxies (LCBGs) (Hoyos \& D\'\i az 2005).

For the upper branch, in the range 12+log(O/H) $\geq$ 8.4 the best fit is found 
with McG91 calibration, although it predicts abundances 0.08 dex larger than the mean value, with
a dispersion of 0.19 dex. Taking into account that the average error in the
direct oxygen abundance determination in this regime is about $\pm$0.20,
this calibration provides a good estimate of the oxygen
abundance for metallic objects.
The calibration by Pil01a underestimates abundances by 0.14 dex on
average with a dispersion of 0.21 dex. 
Contrary to what is found for the lower branch, the only calibration
not taking into account the dependences on log $U$ and T$_{eff}$ (ZKH94) shows
the largest dispersion (0.27 dex) which probably implies that, in
this case, the calibrating sample objects do not share ionising properties. 

Finally, for the intermediate region, in the range 8.0 $\leq$ 12+log(O/H) $<$ 8.4, we have evaluated
both the calibrations for the lower and upper branches. In this case it can be observed that
the lower branch calibrations 
underestimate the metallicity and the upper branch calibrations
overestimate it with residuals that increase with increasing and
decreasing metallicity respectively. In this
regime it is virtually impossible to choose any reliable calibration of
O$_{23}$.

There are not many ways to improve on the O$_{23}$
abundance parameter calibration, since HII regions and HII galaxies are ionised by
young star clusters and, as these clusters evolve, their ionisation
parameters and ionising temperatures change in ways that are not easy
to parametrize. A more promissing approach is the search for other
potentially useful abundance parameters, some of which are examined below.

\subsection{Parameters involving [NII]}

\begin{figure}
\begin{minipage}{85mm}
\psfig{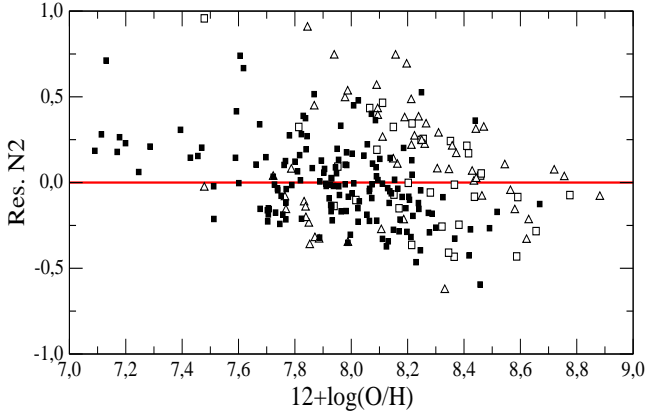}
\caption{The residuals of the fit by the Denicol\'o et al.(2000) calibration 
and the directly derived abundances as a function of the oxygen 
abundance for the compiled data}
\label{resN2}
\end{minipage}
\end{figure} 

\begin{figure}
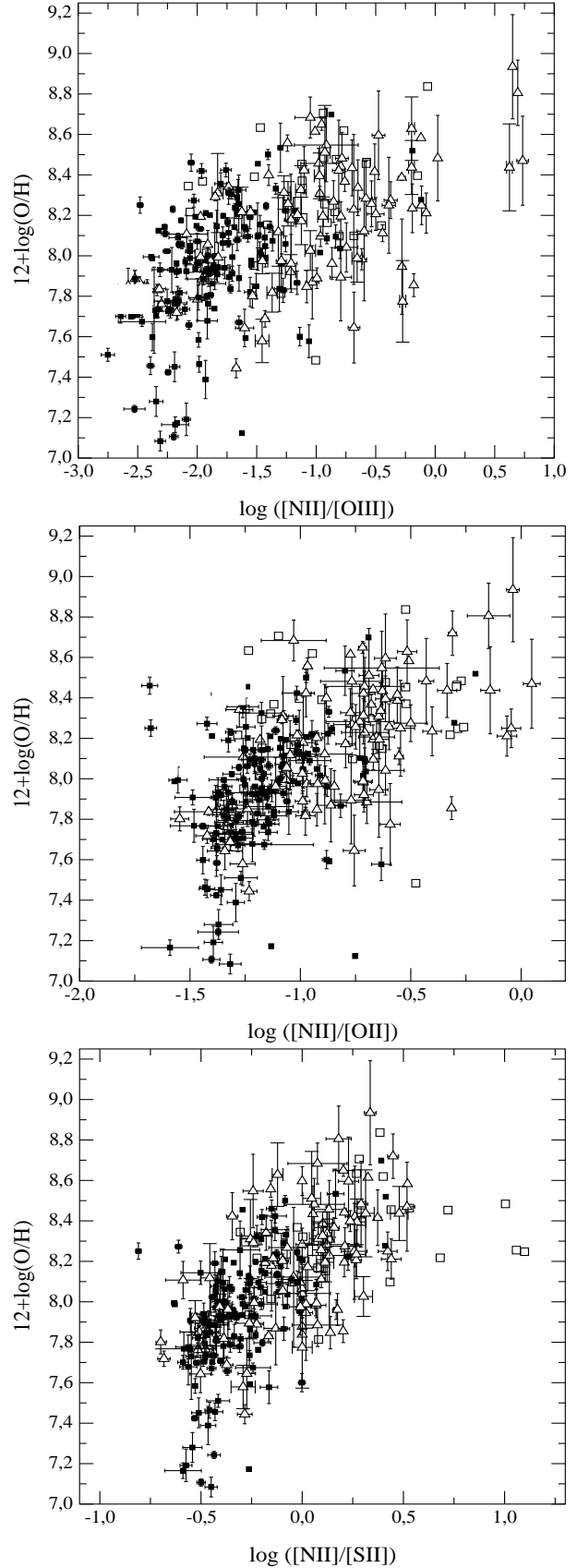

\begin{minipage}{85mm}
\psfig{figure=fig09u.eps,height=7.5cm,width=8cm,clip=}
\psfig{figure=fig09m.eps,height=7.5cm,width=8cm,clip=}
\psfig{figure=fig09d.eps,height=7.5cm,width=8cm,clip=}
\caption{Relation between the ([NII]/[OIII]) (top panel), ([NII]/[OII]) (middle panel) 
and ([NII]/[SII]) (bottom panel) ratios and the metallicity, represented by 12+log(O/H).} 
\label{n2o3_o}
\end{minipage}
\end{figure} 

The N2 parameter was defined as:
\[N2 \equiv \log \frac{I(6584 {\mathrm \AA})}{I(H\alpha)}\]
\noindent by Denicol\'o, Terlevich \& Terlevich (2002, hereinafter DTT02), although it was used 
before as an empirical estimator by Storchi-Bergmann et al. (1994) and
Van Zee et al. (1998). The relation between N2 and the logarithmic
oxygen abundance is shown in the left panel of Figure \ref{N2_o} for all
the objects in the sample for which nitrogen data exists. The
N2 parameter has several advantages: first of all, contrary to O$_{23}$, the relationship between N2
and oxygen abundance is single-valued and secondly, since the emission lines on which it 
is based are very close in wavelength, the N2 parameter is almost free of uncertainties 
introduced by reddening corrections or flux calibrations.
The dash-dotted line in the plot corresponds to the relation found by
DTT02, for their sample objects:
\[12+\log(O/H) = 9.12 + 0.73{\mathrm N2}\]
This line represents a reasonable fit to the data, but shows a large
scatter at all metallicities. Most of the scatter is shown 
by GEHR data, while HII galaxies define a much narrower relation. Most GEHR data fall below the line. This is not surprising since many of the metal rich GEHR used by DTT02 in their calibration had oxygen abundances derived  from  O$ _{23} $ and directly derived abundances for metal rich HII regions tend to be lower than those derived from this parameter (Castellanos et al. 2002; Garnett et al. 2004). 

Again, the main reason
for the dispersion is probably related to the different ionisation parameters and stellar effective temperatures of GEHR since the
N2 parameter depends on both as can be seen in the right panel of Figure \ref{N2_o}. The [NII] lines become weaker as the 
excitation degree and/or
the ionising temperature increase. N2 reaches a maximum of about -0.5 for models
with low effective temperature (35000 K) while the lowest  values of the
parameter are found in models with high effective temperature (50000 K)
and high ionisation parameter (logU=-2.0). An additional source of
scatter is related to the possibly different N/O relative abundances. To try to quantify
this effect we have added a set of photo-ionisation models with a 
value of log(N/O) 0.5 dex lower than our solar assumed value (see Table
\ref{abun}) and log $U = 10^{-2.5}$.  This model sequence can be seen also in Figure \ref{N2_o}
(right panel). A lower N/O ratio mimics a higher ionisation
parameter. Model sequences of high ionising temperatures and constant
N/O ratio seem to reproduce adequately HII galaxy data while a sequence of
models with N/O incresing with oxygen abundance would seem more
adequate for GEHR data. 

The residuals of the fit from the DTT02 calibration against the
directly determined oxygen abundances 
are represented in Figure \ref{resN2}. This empirical calibrator 
works reasonably well in the turnover region of the logO$_{23}$ versus 12+log(O/H) plot, although the  
dispersion reaches a value of 0.27 dex for the reasons given above. This dispersion is of 0.25 dex for HII galaxies. For the rest of the
sample it reaches 0.3 dex with the mean value  0.1 dex higher thus implying that, in these cases, oxygen abundances could be overestimated.

\begin{figure}
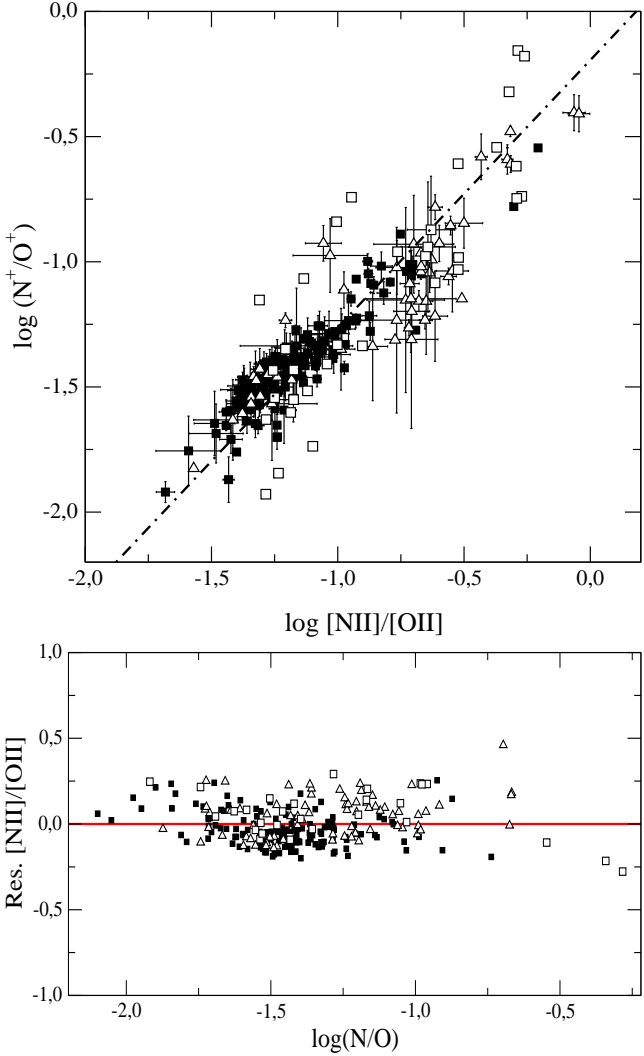

\begin{minipage}{85mm}
\psfig{figure=fig10u.eps,height=8.5cm,width=8.5cm,clip=}
\psfig{figure=fig10d.eps,height=5.5cm,width=8.5cm,clip=}
\caption{Upper panel: Relation between the  ([NII]/[OII]) ratio and the (N$^+$/O$^+$) value for the objects of the sample. 
The dotted-dashed line shows the empirical calibration deduced. Lower
panel: the residuals of the fit as a function of the N/O ratio.} 
\label{n2o2_o}
\end{minipage}
\end{figure} 

Other empirical parameters involving the [NII] lines are the
[NII]$\lambda\lambda$6548,6584 \AA/[OIII]$\lambda\lambda$4959,5007 \AA\ ratio, first proposed by Alloin et al. (1979)
and recently revindicated by Pettini \& Pagel (2004),
and the [NII]$\lambda\lambda$6548,6584 \AA/[OII]$\lambda\lambda$3727,3729 \AA\ and
[NII]$\lambda\lambda$ 6548,6584 \AA/[SII]$\lambda\lambda$ 6716,6731 \AA{} ratios suggested by Dopita \& Evans
(1986) and Kewley \& Dopita (2002) 
as metallicity calibrators in the high abundance regime. In Figure \ref{n2o3_o}, 
12+log(O/H) is
represented as a function of the [NII]/[OIII] parameter (top panel),
the  [NII]/[OII] parameter (middle panel) 
and the [NII]/[SII] parameter (bottom panel) for the compiled sample of objects. 
>From the direct comparison 
with observational data, it can be seen that these three parameters are valid only for a 
metallicity higher than 12+log(O/H) $\approx$ 7.8 and with a scatter
similar to that found for the N2 parameter. This scatter could again be 
related to the different objects presenting different N/O ratios. In
fact, a clear segregation is
found between HII galaxy and GEHR data, more evident in the
upper panel, which is probably related to the GEHR showing higher values of
N/O and hence N/S ratios. Figure
\ref{n2o2_o} (upper panel) shows that a tight relation exists between log([NII]/[OII])
and  log(N$^+$/O$^+$) that is, in turn, a very good indicator of log(N/O). The fit of a regresion 
line to the data produces the relation:
\[\log\left(\frac{N}{O}\right) = 1.144\log\left(\frac{[NII]}{[OII]}\right)-0.232\]
The uncertainty involved in the determination of the
N/O ratio from [NII]/[OII] for the whole sample, represented by the 
standard deviation of the residuals, is 0.14 dex, but decreases to only 0.08 for HII galaxies (see Figure \ref{n2o2_o}, lower panel).

\subsection{The S$_{23(4)}$ parameter}

\begin{figure*}
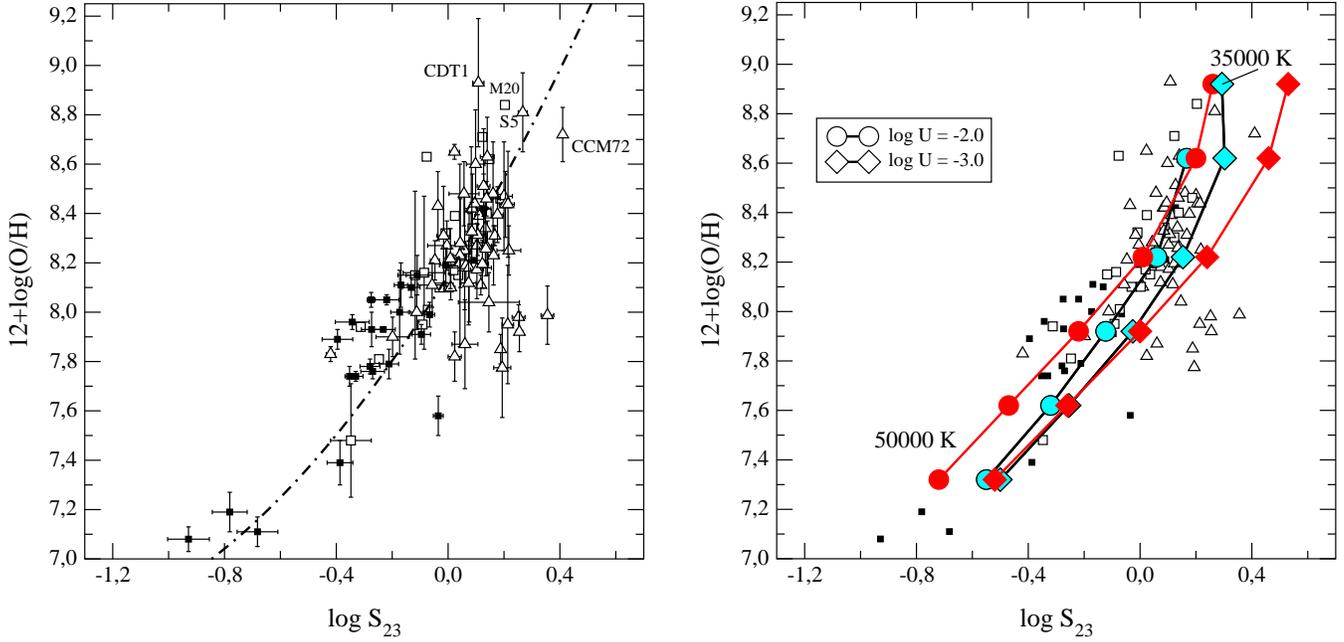

\begin{minipage}{180mm}
\centerline{
\psfig{figure=fig11l.eps,height=8.5cm,width=8.5cm,clip=}
\hspace{0.5cm}
\psfig{figure=fig11r.eps,height=8.5cm,width=8.5cm,clip=}}
\caption{Relation between log $S_{23}$ and the metallicity, represented by 12+log(O/H) ( left) 
and the comparison with CLOUDY photo-ionisation models ( right) for 
different values of effective temperature (35000 K in light tone and 50000 K in dark tone), 
metallicity (from 0.08 to 1.6Z$_{\odot}$) 
and ionisation parameter ($\log U$ = -2.0 (circles) and -3.0 (diamonds). The dashed-dotted line 
shows the calibration derived in this work.}
\label{s23_o}
\end{minipage}
\end{figure*}

The S$_{23}$ parameter was defined by V\'{\i}lchez \& Esteban (1996) as: \\
\[S_{23} \equiv \frac {{\mathrm I(6717\AA)+I(6731\AA)+I(9069\AA)+I(9532\AA)}}{{\mathrm I(H}\beta)}\]
using the [SII] and [SIII] lines analogous to those of [OII] and [OIII]
in the O$_{23}$ parameter. It was proposed by Christensen, Petersen \&
Gammerlgaard (1997) as a sulphur abundance indicator and more recently
by DPM00 as an oxygen abundance indicator due to the characteristics evident in
Figure \ref{s23_o} (left panel): firstly its single-valued 
behaviour up to solar metallicities, and secondly its lower dependence
on the other functional parameters. The fact that S$_{23}$ presents a lesser
dependence on effective temperature and ionisation parameter than
$O_{23}$, seems to be confirmed by photo-ionisation models (see Figure \ref{s23_o}, right
panel). There is a dependence of S$_{23}$ on log$U$ and T$_{eff}$ but
it is weaker than for O$_{23}$ (see Figure \ref{o23_o} for a
comparison).

>From the observational point of view, the S$_{23}$ parameter has two
important advantages: firstly the sulphur lines remain intense even for
the highest metallicity objects and secondly it is relatively 
independent of reddening, since the lines of both [SII] and [SIII] can be
measured relative to nearby hydrogen recombination lines. On the
negative side, [SIII] lines shift out of the far red spectral region
for redshifts higher than 0.1.

\begin{figure*}
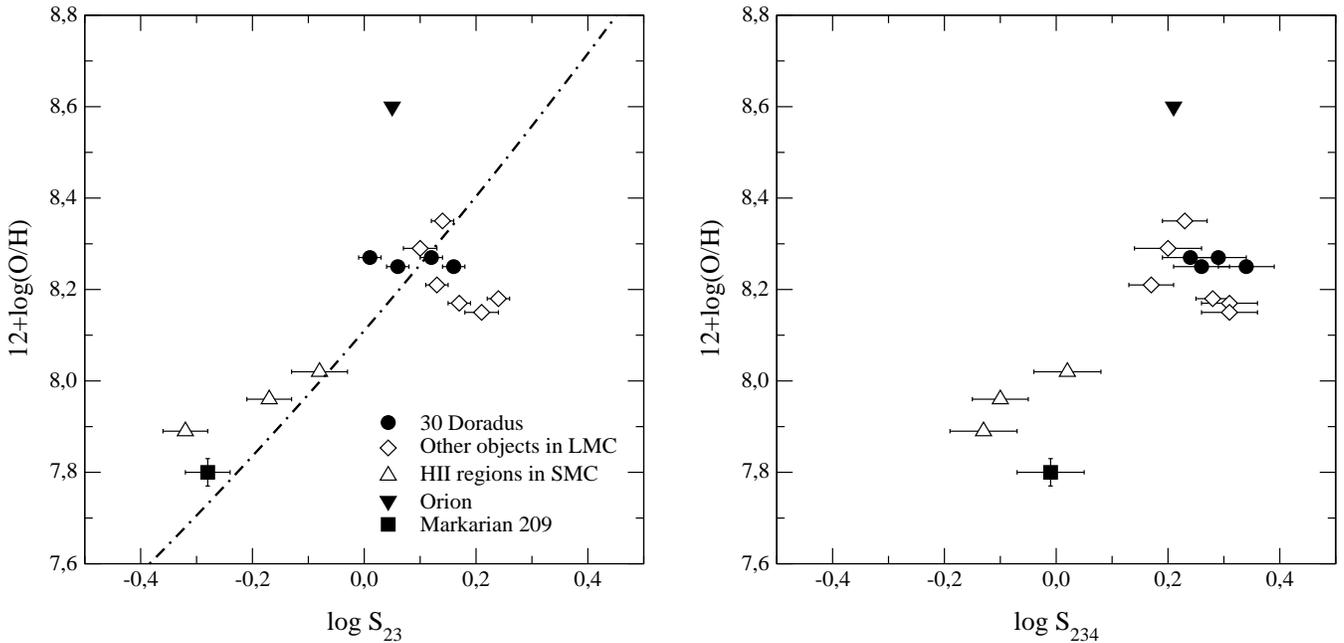

\begin{minipage}{180mm}
\centerline{
\psfig{figure=fig12l.eps,height=8.5cm,width=8.5cm,clip=}
\hspace{0.5cm}
\psfig{figure=fig12r.eps,height=8.5cm,width=8.5cm,clip=}}
\caption{Relation between logS$_{23}$ (left) and logS$_{234}$ (right) and the oxygen abundance 
for the sample of objects with a direct measurement of the [SIV] 10.52$\mu$ line: Mrk 209 (filled square), 
Orion (filled triangle), 30 Doradus (filled 
circles) and objects from LMC (open diamonds) and SMC (open triangles).}
\label{s234_o}
\end{minipage}
\end{figure*} 
Using the newly added observational 
data we have improved the DPM00 relation to:
\[12+log(O/H) = 8.15+1.85\log S_{23}+0.58(\log S_{23})^2\]
whose residuals for the complete sample relative to the directly
determined oxygen abundances are represented as a function of oxygen
abundance in the upper panel of Figure \ref{ResS23} . The dispersion is approximately
equal to 0.2 dex in all the abundance ranges although it decreases to
0.10 dex for the HII galaxy sample. 
The relation is not linear.  Values of S$_{23}$ lower than expected are found for 
higher excitation nebulae having low metallicity probably due to 
the presence of [SIV] in non-negligible amounts, 
as seems to be indicated by the position on the diagram 
of IZw18, the least metallic object. Unfortunately, despite recent 
observations of a sample of HII galaxies in the near IR spectral range (PMD03), there is 
a considerable lack of data on objects of low metallicity  whose
inclusion would definitely improve the calibration.

\begin{figure}
\begin{minipage}{85mm}
\psfig{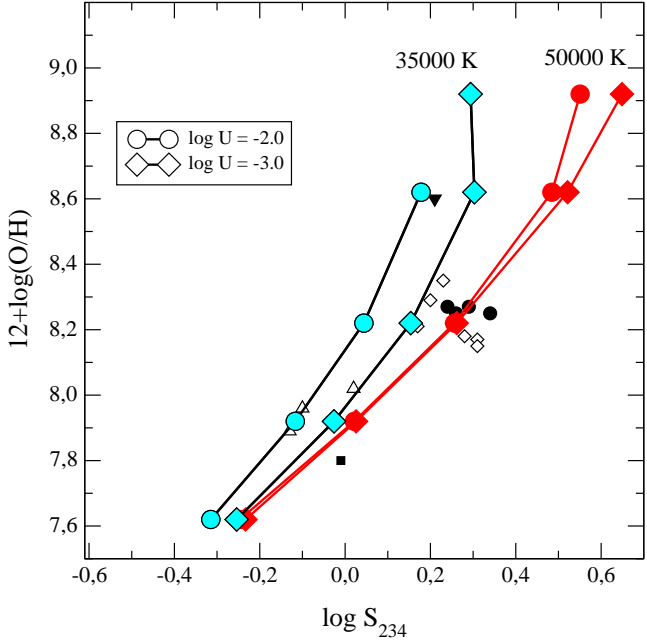}
\caption{Comparison between observations and model results for the 
relation between the S$_{234}$ parameter and the oxygen abundance.}
\label{ModS234}
\end{minipage}
\end{figure}

Oey \& Shields (2000) have defined a new parameter, S$_{234}$, 
\[S_{234}\equiv\frac{{\mathrm I(6725\AA)+I(9069\AA)+I(9532\AA)+I(10.5}\mu)}{{\mathrm I(H}\beta)}\]
which takes into account the contribution of [SIV] 
through its emission line at 10.52 $\mu$. 
The contribution of [SIV] is expected to be relevant only in objects
with a high degree of ionisation (D\'{\i}az et al. 1991) and therefore the use of S$_{234}$
almost eliminates the dependence on ionisation parameter found for
S$_{23}$ (Kennicutt et al. 2000). In fact, photoinisation
models indicate that S$_{23}$ is only slightly
dependent on ionisation parameter but shows a non negligible dependence on effective 
temperature, which becomes more evident at high metallicities (see Figure
\ref{ModS234}).

 Unfortunately, the sample of objects for which the [SIV] line 
in the mid infrared is measured is very poor. Using the available data
for these objects we have confirmed that the contribution of this
line to S$_{234}$ can be rather large. We have found very little [SIV]
data for HII regions, GEHR and HII galaxies: the Orion nebula (Lester
et al. 1979),  Mrk 209 (Nollenberg et al. 2002) and a sample of
objects in the Magellanic Clouds 
(Vermeij et al. 2002). Data on these objects are plotted in Figure
\ref{s234_o} (S$_{23}$ in the left panel and S$_{234}$ in the right panel). 
Any improvement in the abundance calibration is difficult to quantify
given the scarcity of data. 
At any rate, since no observations
of the $\lambda$ 10.5 $\mu$ line exist for most objects, it would have to be
calculated from photo-ionisation models which would make  S$_{234}$ a
semi-empirical parameter.

\subsection{The S$_{23}$/O$_{23}$ parameter}

One fact that becomes evident from the examination of the different
abundance parameters discussed above is that the validity of each one of them seems to be
restricted to a given metallicity range. This means that it is
necessary to have some {\em a priori} knowledge about the metallicity of an object or
a sample of objects in order to choose the appropriate abundance
indicator.  

Traditional ways of doing this include an examination of the [NII]$\lambda$
6584 \AA/H$\alpha$ ratio that can discriminate between objects with
12+log(O/H) higher or lower than about 8.0 (S89). More recently, the values of
the O$_{23}$ and S$_{23}$ taken together have been used by DPM00 to discriminate between
subsolar and oversolar abundances.
However, in some cases, the interest is focused on the comparison of global 
abundances trends shown by different objects  
or abundance distributions over a wide range of metallicities. In those
cases it would be desirable, to obtain abundances by means of the same
calibrator so that comparisons are meaningful. This calibrator should
be valid for the whole metallicity range.

\begin{figure}
\begin{minipage}{85mm}
\psfig{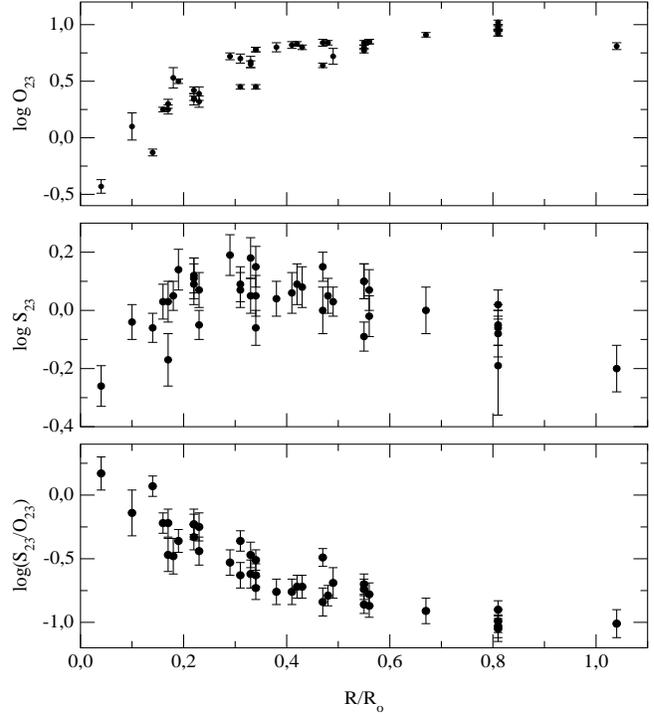}
\caption{Variation of log O$_{23}$ (top panel), log S$_{23}$ (middle panel) and log (S$_{23}$/O$_{23}$) 
(bottom panel) with normalised galactocentric distance for M101 (data from Kennicutt \& Garnett 1996).}
\label{m101}
\end{minipage}
\end{figure} 

The study of metallicity gradients over galaxy discs is one of the issues what could be improved 
in this way. For example, there are many different conclusions  
about the value of the oxygen abundance distribution in the well studied galaxy M101 . Different authors 
(Zaritsky, 1992; Scowen et al. 1992; Vila-Costas \& Edmunds, 1992) have pointed to an increase in the 
slope of the gradient in the inner regions whereas other authors (Henry \& Howard, 1995; Pilyugin, 2001b) 
obtain an exponential law throughout the whole disc. 
Kennicutt \& Garnett (1996) 
have shown how the use of one calibration of O$_{23}$ or another leads to different conclusions.

\begin{figure*}
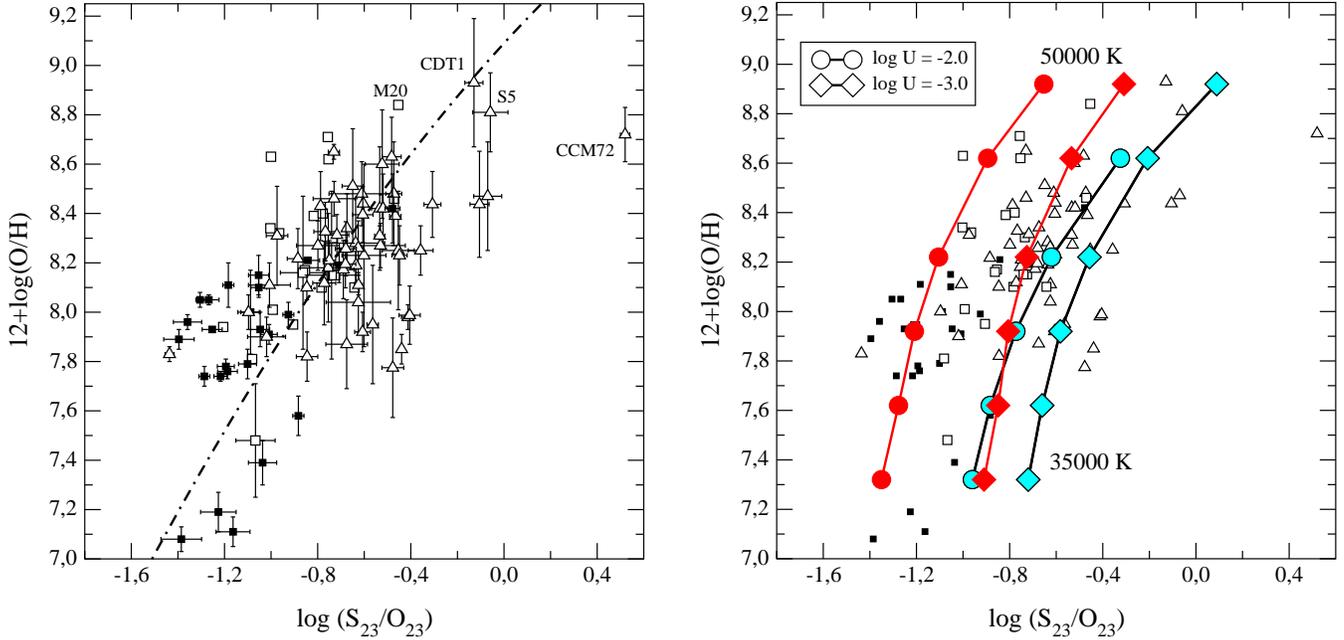

\begin{minipage}{180mm}
\centerline{
\psfig{figure=fig15l.eps,height=8.5cm,width=8.5cm,clip=}
\hspace{0.5cm}
\psfig{figure=fig15r.eps,height=8.5cm,width=8.5cm,clip=}}
\caption{Relation between log ($S_{23}/O_{23}$) and the metallicity, represented by 12+log(O/H) (at left) 
and the comparison with CLOUDY photo-ionisation models  (right) for 
different values of effective temperature (35000 K in light tone and 50000 K in dark tone), 
metallicity (from 0.08 to 1.6Z$_{\odot}$)
and ionisation parameter ($\log U$ = -2.0 (circles) and -3.0 (diamonds). The dashed-dotted line corresponds to the calibration proposed in this work.}
\label{s23o23_o}
\end{minipage}
\end{figure*} 

In Figure \ref{m101} we 
represent the gradient of some of the parameters studied here as a function of the galactocentric 
distance to the center of M101 (data from Kennicutt \& Garnett, 1996). In the upper panel the
O$_{23}$ parameter is seen to increase with increasing galactocentric radius up to a value of
0.3 R$_0$ and then it remains almost constant. In the middle panel, the S$_{23}$ parameter shows the
same behaviour as O$_{23}$ in the central regions of the disc but decreases with increasing
galactocentric radius from 0.3 R$_0$ onwards. These two trends taken together point to the central
disc regions of M101 ( R $<$ 0.3 R$_0$ ) being over-solar and thus lying on the upper branch of the O$_{23}$ and
S$_{23}$ parameters. The outer regions of the disc would have under-solar abundances and lie on the
lower branch of the S$_{23}$ calibration. Most of the regions in this regime ( R $>$ 0.3 R$_0$ )  lie 
on the turnover region of the O$_{23}$ calibration and therefore show an almost constant value of this parameter. 
In the lower panel of Figure \ref{m101}, 
we can see that a combination of the two parameters, S$_{23}$/O$_{23}$, shows a continously
decreasing trend through the disc of M101. 

Using all the objects of our sample with measurements of the [OII], [OIII], [SII] and [SIII] lines and a direct 
determination of the metallicity, we have calibrated for the first time this new parameter (Figure 
\ref{s23o23_o} left panel). As can be seen, the relation remains single-valued for the whole 
range of metallicity, though it is non-linear, due probably to the contribution 
of [SIV] for high excitation regions of low metallicity. The addition of high 
metallicity objects is not possible due to the lack of data with he necessary auroral lines, but the 
fact that the parameter increases towards the 
inner parts of the disc of M101 (see Figure \ref{m101}) suggests that it does not undergo any 
turnover at high metallicites. Of course, this parameter keeps the same sources of uncertainty 
as its two progenitors. In Figure \ref{s23o23_o} (right), where we compare the observational data 
with results from our photo-ionisation models, it can be seen that the value of S$_{23}$/O$_{23}$ 
increases for low degrees of ionisation and lower effective temperatures.

\begin{figure}
\begin{minipage}{85mm}
\psfig{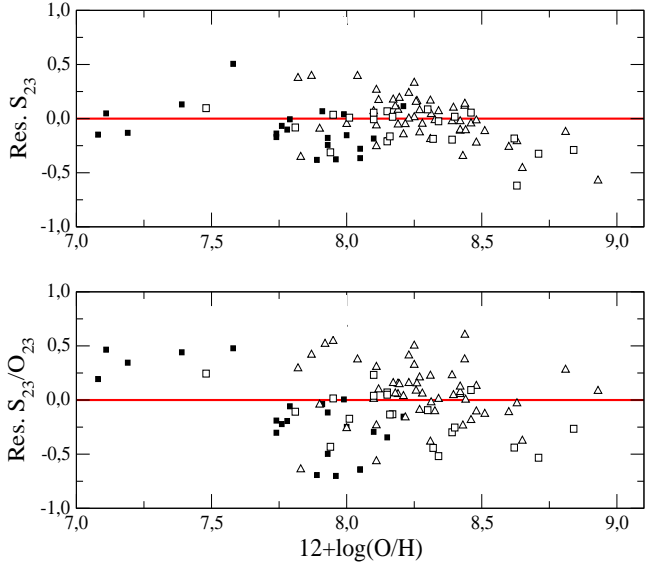}
\caption{Residuals of the abundances determined from the calibration with S$_{23}$ (upper panel) and
S$_{23}$/O$_{23}$ (bottom panel) 
to those derived directly, as a function of abundance.}
\label{ResS23}
\end{minipage}
\end{figure}

Using the compiled data, 
we propose the following relation to derive oxygen abundances in the whole metallicity range in the 
absence of auroral emission line data:
\[12+\log(O/H)=9.09+1.03\log\left(\frac{S_{23}}{O_{23}}\right)-0.23\left[\log\left(\frac{S_{23}}{O_{23}}\right)\right]^2\]
This relation is plotted along with data in the left panel of Figure \ref{s23o23_o}. The residuals of this fit from
 the values of 12+log(O/H) deduced from the direct method, are plotted in the lower panel of Figure \ref{ResS23} 
and show a dispersion of 0.27 dex, comparable to that found for the N2 parameter.

\begin{figure}

\begin{minipage}{85mm}
\psfig{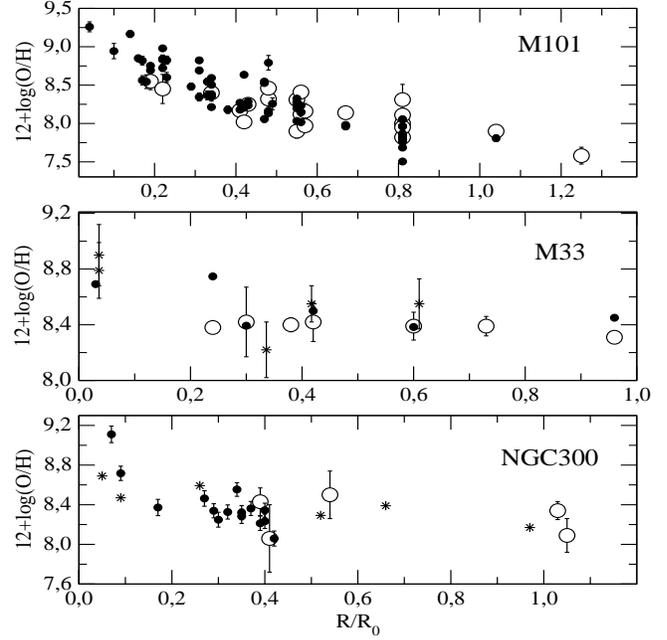}
\caption{Metallicity gradients of M101, M33 and NGC300. Filled circles: abundances derived from the 
S$_{23}$/O$_{23}$ parameter; open circles: abundances derived from the direct method; asterisks: 
abundances derived for early type stars (see text for details).}
\label{gradients}
\end{minipage}
\end{figure}

The upper panel of Figure \ref{gradients} shows the oxygen abundance gradient in M101. Solid circles 
correspond to oxygen abundances derived using the S$_{23}$/O$_{23}$ parameter from the emission line data 
of Kennicutt \& Garnett (1996); open circles correspond to abundances derived by the direct method 
(Kennicutt et al. 2003). The agreement between directly determined and empirically derived abundances is excellent. 

Unfortunately, sulphur line intensity data is scarce and therefore it is difficult to assess the suitability of the
S$_{23}$/O$_{23}$ parameter as an abundance gradient indicator. We have found only two more galaxies
with data covering a substantial part of the disc: M33 (data from Kwitter \& Aller 1981 and
V\'\i lchez et al. 1988) and NGC300 (data from Deharveng et al. 1988 and Christensen et al. 1997).
Their abundance gradients are shown in the middle and lower panels of Figure \ref{gradients} respectively. Again, solid circles 
correspond to oxygen abundances derived using the S$_{23}$/O$_{23}$ parameter and open circles correspond to 
abundances derived by the direct method. We have also added oxygen abundance data derived from the spectroscopic 
analysis of  early B-type supergiant stars (Monteverde et al. 1988; Urbaneja et al. 2003). 
Although the discs of these two galaxies are not as well sampled as that of M101, the agreement 
between empirically and directly derived abundances is good and the agreement between nebular and 
stellar abundances is encouraging.  Regarding the shape of the gradients, in all three cases an increase 
in slope for the central galactic regions is apparent, although in the cases of M33 and NGC300 nothing 
conclusive can be said. We think that more observations and more work along these lines would greatly 
help to derive the true abundance distributions across the discs of galaxies.  

\section{Summary and conclusions}

In this work we have revised the different proposed oxygen abundance calibrations using a large compiled sample
of observations comprising the emission lines of [OII], [OIII], [SII], [SIII] and [NII] for objects
with oxygen abundances derived by the direct method. The data has been compared with
results from a set of photo-ionisation models in order to seek an explanation to
the sources of scatter in the calibrations. The direct calibration of any parameter with model results alone
would lead to non quantifiable systematic errors and we consider  strictly empirical calibrations to be  much
more reliable.


\begin{table}
\begin{minipage}{85mm}
\footnotesize
\caption{Summary of the properties of the different empirical calibrators. The
uncertainty is found from the standard deviation of the residuals of each parameter
with the oxygen abundance derived directly. See the text for the mean deviations.}
\begin{center}
\begin{tabular}{lcc}
\hline
\hline
Parameter & Range of Z \footnote{Z represents 12+log(O/H)} & 
Uncertainty\footnote{in logarithmic units, dex} \\
\hline
O$_{23}$ (lower branch) & Z $<$ 8.0 & 0.13 (Pil00, McG91) \\ 
O$_{23}$ (int. region) & 8.0 $\leq$ Z $<$ 8.4 & $\approx$ 0.70 \\ 
O$_{23}$ (upper branch) & Z $\geq$ 8.4 & 0.19 (McG91) \\
N2  & All Z (HII galaxies) & 0.25 \\
S$_{23}$ & Until 8.9 & 0.20 \\
S$_{23}$/O$_{23}$ & All Z & 0.27 \\  
\hline
\end{tabular}
\end{center}
\label{summ}
\end{minipage}
\end{table}

In Table \ref{summ} we summarize the main properties of the parameters studied, including 
their metallicity range of validity, and the 
uncertainty obtained for each calibration, understood as the standard deviation of the residuals of 
the deduced oxygen abundance and that derived through the direct method. 
The O$_{23}$ parameter is 
the most widely used due to the important role of oxygen in the cooling of the ionised gas and
because the oxygen emission lines remain in the optical-far red  part of the spectrum until
redshift $\approx$ 1. This parameter presents a 
double-valued relation with metallicity and in the lower and upper branch (in the latter case,
taking into account the strong dependence on effective temperature) the uncertainty remains below
 0.2 dex. Nevertheless, the uncertainty in the turnover region (for 12+log(O/H) between
8.0 and 8.4) may reach almost an order of magnitude. No ways of improving this calibration further have been found in this work.

The best alternative to this parameter in this metallicity regime is S$_{23}$. 
In objects where it is possible to observe the near-IR [SIII] lines (up to redshift $\approx$ 0.1), 
the metallicity can be deduced with less than 0.2 dex dispersion up to oxygen abundances  12+log(O/H) $\approx$ 8.9.
The S$_{23}$ parameter also offers the advantage of being relatively independent 
of reddening. 
 The contribution of the [SIV] emission line in the mid-IR is relevant only for high excitation
objects and can be taken into account by means of the S$_{234}$ parameter. Unfortunately, there 
is, at the moment, very little data to calibrate it empirically, and avalaible photo-ionisation models fail 
to correctly reproduce the ionisation structure for sulphur (PMD03). 

The N2 parameter, also reddening independent,  is a good alternative for distant objects (up to $z$ $\approx$ 0.5)
with intermediate metallicity ( 8.0 $\leq$ 12+log(O/H) $\leq$ 8.4). Nevertheless, 
this parameter suffers from uncertainties due to its dependence on ionisation parameter and N/O ratio.
Besides, the calibration done by DTT02 works better for HII galaxies data but its application to
higher metallicity regions carries a higher uncertainty and possibly overestimates the derived
oxygen abundances.

Regarding the [NII]/[OIII], [NII]/[OII] and [NII]/[SII] parameters, these are only correlated with oxygen abundances 
at moderate to high metallicities and show an uncertainty similar to that of N2. However, as the main source of uncertainty for the calibrations involving the nitrogen lines 
is probably related to the N/O ratio, these calibrations could be improved with the use of [NII]/[OII] as a N/O calibrator which shows a 
dispersion of only 0.10 dex for HII galaxies.

Finally we have used the compiled sample of objects to produce for the first time a calibration
for the S$_{23}$/O$_{23}$ parameter,  which could be useful to study variations over a wide rage of metallicities, 
as is the case for the discs of galaxies. This new parameter includes non-negligible uncertanties, inherited 
from its two predecesors, O$_{23}$ and S$_{23}$ but the results for the disc of M101 are encouraging.

\section*{Acknowledgements}

We would like to thank M. Castellanos, R. Terlevich, E. Terlevich, C. Esteban, E. P\'erez and
D. Valls-Gabaud for very interesting discussions and suggestions and an anonimous referee for a careful revision of the manuscript.  We would also like to acknowledge the thorough revision of the English done by Michael Taylor.\\
This work has been partially supported by DGICYT projects AYA-2000-0973
and AYA-2004-08260-C03-03.

\end{document}